\documentclass[pdflatex,sn-mathphys-num]{sn-jnl}

\usepackage{graphicx} 
\usepackage{amsmath,amssymb}
\usepackage{algorithm,algpseudocode}
\usepackage{float}

\begin{document}

\title[Dual-Attention U-Net++ with Class-Specific Ensembles and Bayesian Hyperparameter Optimization for Precise Wound and Scale Marker Segmentation]{Dual-Attention U-Net++ with Class-Specific Ensembles and Bayesian Hyperparameter Optimization for Precise Wound and Scale Marker Segmentation}

\author*[1,2]{\fnm{Daniel} \sur{Cieślak}}\email{daniel.cieslak@pg.edu.pl}

\author[1]{\fnm{Miriam} \sur{Reca}}

\author[1]{\fnm{Olena} \sur{Onyshchenko}}

\author[1]{\fnm{Jacek} \sur{Rumiński}}

\affil*[1]{\orgdiv{Faculty of Electronics, Telecommunications and Informatics}, \orgname{Gdańsk University of Technology}, \city{Gdańsk}, \country{Poland}}

\affil[2]{\orgname{IDEAS NCBR}, \city{Warsaw}, \country{Poland}}

\abstract{
Accurate segmentation of wounds and scale markers in clinical images remains a significant challenge, crucial for effective wound management and automated assessment. In this study, we propose a novel dual-attention U-Net++ architecture, integrating channel-wise (SCSE) and spatial attention mechanisms to address severe class imbalance and variability in medical images effectively. Initially, extensive benchmarking across diverse architectures and encoders via 5-fold cross-validation identified EfficientNet-B7 as the optimal encoder backbone. Subsequently, we independently trained two class-specific models with tailored preprocessing, extensive data augmentation, and Bayesian hyperparameter tuning (WandB sweeps). The final model ensemble utilized Test Time Augmentation to further enhance prediction reliability. Our approach was evaluated on a benchmark dataset from the NBC 2025 \& PCBBE 2025 competition. Segmentation performance was quantified using a weighted F1-score (75\% wounds, 25\% scale markers), calculated externally by competition organizers on undisclosed hardware. The proposed approach achieved an F1-score of 0.8640, underscoring its effectiveness for complex medical segmentation tasks.
}


\maketitle

\section{Introduction}

The skin, being a vital organ, is usually categorized as the body's largest. The complex, layered structure acts as a safeguard. Although strong, it is simultaneously sensitive and easily affected by injuries from accidents, extended illnesses, surgeries, and other causes. Proper and precise documentation of wounds is essential for their effective healing.  In addition, detecting skin conditions accurately is vital for reducing incidence, treatment costs, morbidity, and mortality, highlighting the need to examine damaged skin \cite{yazdi2022mechanical}. Medical practitioners have historically relied on manual measurements, photos, and visual assessments to track and evaluate wound healing. While these techniques are generally effective, they are subjective, prone to mistakes, and depend on the healthcare provider's experience \cite{foltynski2023wound}. This variability emphasizes the pressing need for technological progress in wound evaluation.

Recent advancements in computer vision and deep learning present a compelling alternative for manual methods. Automated segmentation methods promise objectivity, consistency, and efficiency by transforming pixels on a medical image into insightful clinical data. However, developing an effective segmentation system isn't straightforward; wounds can be irregularly shaped, vary significantly in appearance, and imaging conditions may differ greatly. Additionally, precise measurements require reliable scale markers within images, further complicating the task.

Our goal for the NBC 2025 \& PCBBE 2025 wound image segmentation challenge was not only to benchmark cutting-edge neural networks but also to contribute to a broader narrative—shifting from manual subjectivity toward accurate, reproducible wound assessments powered by artificial intelligence.

In this paper, we present our approach — from thorough exploration of contemporary deep-learning architectures, through rigorous experimentation, to identifying the most promising strategy for simultaneously segmenting wounds and scale markers. Our findings illustrate the strengths and limitations of current state-of-the-art methods and set the stage for further innovations, bringing us closer to making automated, reliable wound assessments a clinical reality.

\section{Related work}
Classical image processing techniques, such as thresholding, contour models and region segmentation, rely on handcrafted features and minimal labeled data but face challenges with complex textures and lighting variations \cite{li2018composite}. The U-Net architecture \cite{ronneberger2015unet} was initially introduced for medical image segmentation, its main advantage is that it can be trained on limited-size datasets. A dual attention U-Net model was proposed for wound segmentation, integrating VGG16 and U-Net architectures with dual attention mechanisms to focus on relevant regions within the wound area. This model achieved high performance metrics, including a Dice coefficient of 94.1\% and IoU of 89.3\% on test data \cite{niri2025wound}. 

\begin{table}[h]
    \centering
    \small 
    \resizebox{\textwidth}{!}{ 
        \begin{tabular}{|p{2.8cm}|p{4.5cm}|p{4.5cm}|p{2.8cm}|}
            \hline
            \textbf{Architecture} & \textbf{Strengths} & \textbf{Weaknesses} & \textbf{Multi-class performance} \\ \hline
            \raggedright SegFormer \cite{xie2021segformer} & \raggedright Combines transformers with lightweight decoders. Captures global context and fine-grained details efficiently. & \raggedright Requires larger datasets. May underperform on highly imbalanced classes. & \raggedright Excellent for multi-class tasks with sufficient data. \\ \hline
            \raggedright U-Net++ \cite{10.1007/978-3-030-00889-5_1} & \raggedright Nested skip pathways and deep supervision improve accuracy for multi-scale structures. Reduces semantic gaps. & \raggedright Increased model complexity. Higher computational costs. & \raggedright Superior for multi-scale structures like wounds and markers. \\ \hline
            \raggedright DeepLab V3+ \cite{chen2018deeplabv3+} & \raggedright Captures multi-scale contextual information. Handles class imbalance well. & \raggedright Computationally intensive. Less effective on small datasets. & \raggedright Excels in multi-class scenarios with sufficient resources. \\ \hline
            \raggedright U-Net & \raggedright Simple encoder-decoder structure. Reliable with limited data. Widely validated. & \raggedright Limited contextual awareness. Struggles with fine-grained multi-class tasks. & \raggedright Robust baseline but less effective for complex multi-class tasks. \\ \hline
        \end{tabular}
    }
    \caption{Comparison of different neural network architectures for segmentation tasks}
    \label{tab:architecture_comparison}
\end{table}

Class imbalance in wound segmentation datasets leads to biased model performance, with studies reporting mean IoU as low as 48.6\% for minority classes like subcutaneous hematomas, but strategies like dual-encoder architectures (SRU-Net) and loss weighting improve accuracy in multi-class scenarios, achieving reliable classification for 7 common wound types despite imbalance \cite{cui2023gan}. The use of attention mechanisms in wound segmentation models could enhance their ability to focus on relevant features, significantly improving accuracy by selectively weighting spatial and channel-wise information, as demonstrated in dual attention U-Nets achieving high Dice coefficients in medical image segmentation tasks \cite{zimmermann2023wound}.

While recent dual-attention architectures reported promising Dice coefficients (e.g., 94.1\% by \cite{niri2025wound}), our study further improves segmentation accuracy through the strategic choice of a stronger encoder (EfficientNet-B7), rigorous Bayesian hyperparameter tuning, and dedicated class-specific ensembles, directly addressing the severe class imbalance and variability present in clinical images

\section{Methods}\label{sec:methods}

\subsection{Dataset}
We utilized two distinct datasets to ensure robust training and evaluation of our segmentation models. The primary dataset, provided by the NBC 2025 \& PCBBE 2025 challenge organizers, contained images annotated with wounds and scale markers. 

Specifically, the training set comprised 371 images in total, of which 208 had two labels (annotated with wounds and scale markers) and 163 without second label. The validation set consisted of 80 images, with 79 annotated with 2 labels and only 1 labeled with one class, clearly highlighting the imbalance issue we aimed to address in our methodological approach.

The secondary dataset, sourced from MICCAI 2020 \cite{info15030140}, included only wound annotations, serving as an auxiliary dataset for pretraining.

Due to significant class imbalance, we trained separate models for wounds and scale markers, each optimized individually with customized preprocessing.

\subsection{Image Preprocessing}
We optimized preprocessing strategies separately for each class, employing intensive data augmentations, including flips, rotations, brightness-contrast adjustments, and geometric distortions. Images were normalized based on ImageNet statistics. Augmentation strength was optimized through Bayesian sweeps (WandB), effectively enhancing generalization.

Augmentations were applied to 100\% of our training images using the Albumentations library. Specifically, these included horizontal and vertical flips (probability 50\%), random rotations by multiples of 90 degrees (probability 50\%), random brightness adjustments (limits $[-0.2274, +0.4548]$) and contrast adjustments ($\pm0.2274$, probability 80\%), geometric distortions such as grid distortion (limit $\pm0.3411$, probability 50\%), and grid dropout (ratio 0.3, probability 50\%). The augmentation strength was optimized through Bayesian sweeps using WandB, with the final strength parameter set to 1.137.

\subsection{Segmentation Architectures}
\begin{figure}[ht!]
    \centering
    \includegraphics[width=0.8\textwidth]{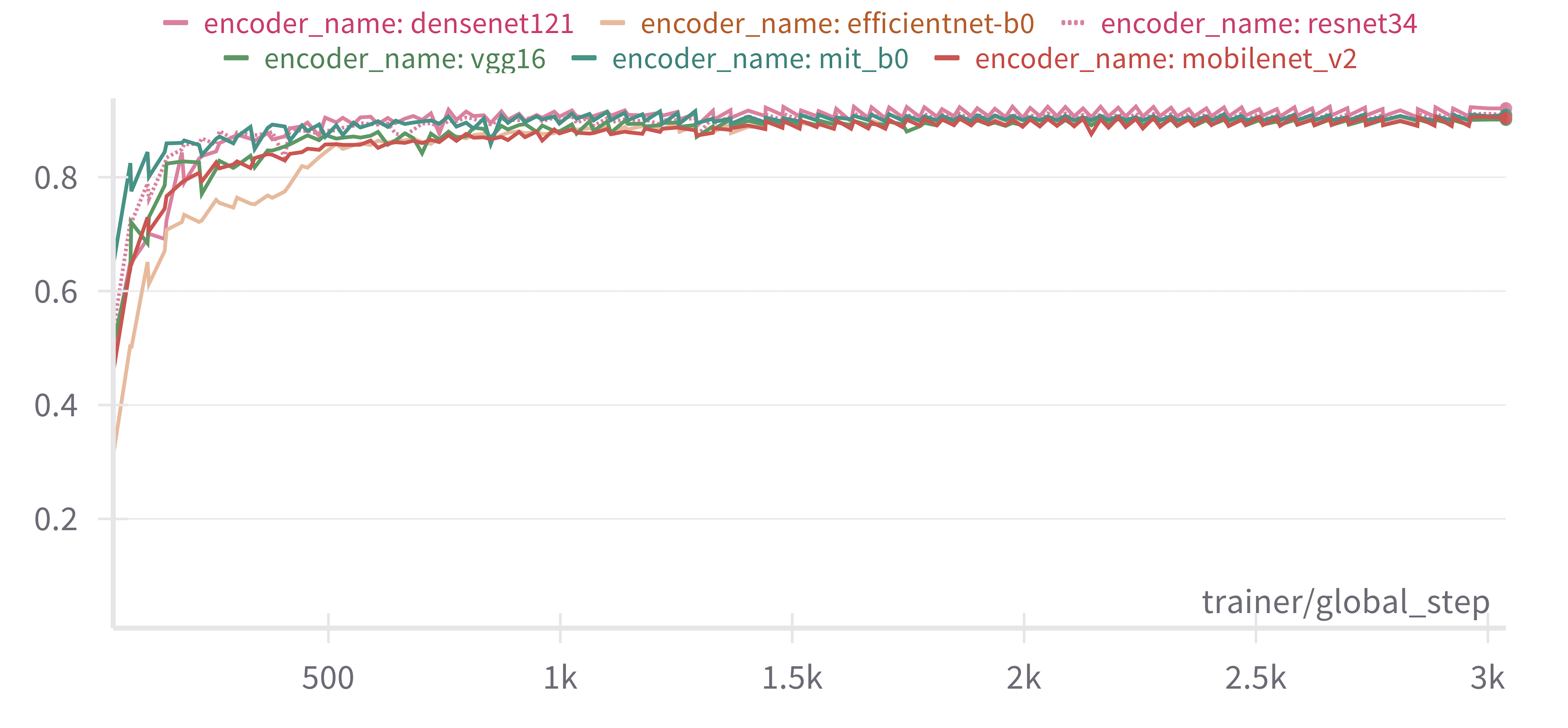}
    \caption{Encoder backbone benchmarking results (validation Dice scores).}
    \label{fig:encoders}
\end{figure}

\begin{figure}[ht!]
    \centering
    \includegraphics[width=0.8\textwidth]{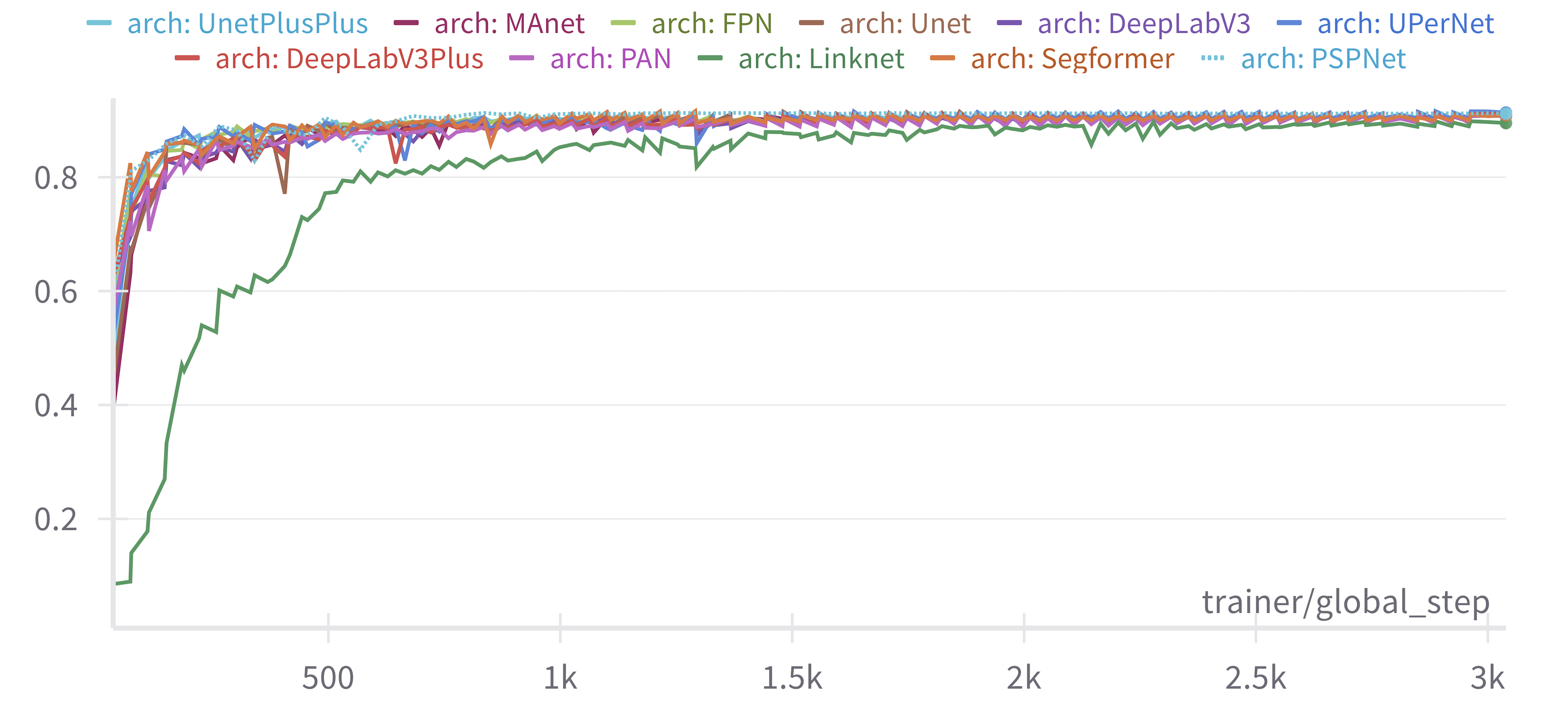}
    \caption{Performance comparison of segmentation architectures (EfficientNet-B0 encoder).}
    \label{fig:architectures}
\end{figure}

Initially, extensive benchmarking was conducted using Segmentation Models PyTorch (SMP) library architectures: U-Net, U-Net++, MAnet, LinkNet, FPN, PSPNet, PAN, DeepLabV3, DeepLabV3+, UPerNet, and SegFormer. These architectures were evaluated across multiple encoders (DenseNet121, EfficientNet-B0, ResNet34, VGG16, MiT-B0, and MobileNetV2) using 5-fold cross-validation on NVIDIA A5000 GPUs. Benchmark results (Fig.~\ref{fig:encoders}, Fig.~\ref{fig:architectures}) identified U-Net++ with EfficientNet-B7 encoder as superior.

We further enhanced U-Net++ by integrating a dual-attention module: spatial attention and SCSE channel-wise attention.

\subsection{Training Protocol}
Models for wounds and markers were trained independently due to data imbalance.

This approach effectively mitigated class imbalance, enabling targeted optimization for each specific class. Separate training allowed easier prototyping and rapid introduction of corrective measures for particularly challenging segmentation cases, significantly improving the overall robustness and performance of the solution.

Both utilized dual-attention U-Net++ with EfficientNet-B7 encoder, optimized via Bayesian hyperparameter sweeps on WandB (loss ratios, augmentation strength, learning rate, mixup alpha).

We trained models using PyTorch Lightning, distributed across two NVIDIA A5000 GPUs, with mixed-precision training. Training ran for 80 epochs, monitored via validation Dice.

Test Time Augmentation aggregated predictions from rotations (0°, 90°, 180°, 270°). Conditional Random Fields were optionally applied post-inference. Final performance was averaged across five runs.

\subsection{Implementation Details}
Key hyperparameters (optimized via Bayesian hyperparameter sweeps using WandB) included:
\begin{itemize}
    \item Architecture: U-Net++ (dual-attention: SCSE + Spatial)
    \item Encoder: EfficientNet-B7 (ImageNet pretrained)
    \item Optimizer: AdamW (lr=0.001; ReduceLROnPlateau scheduler)
    \item Loss ratios optimized via WandB
    \item Batch size: 4; Epochs: 80
    \item Mixup alpha: 0.3604; Augmentation strength: 1.137
    \item Precision: Mixed (FP16)
\end{itemize}

\subsection{Overall Framework Structure}

\begin{figure}[H]
    \centering
    \includegraphics[height=0.8\textwidth]{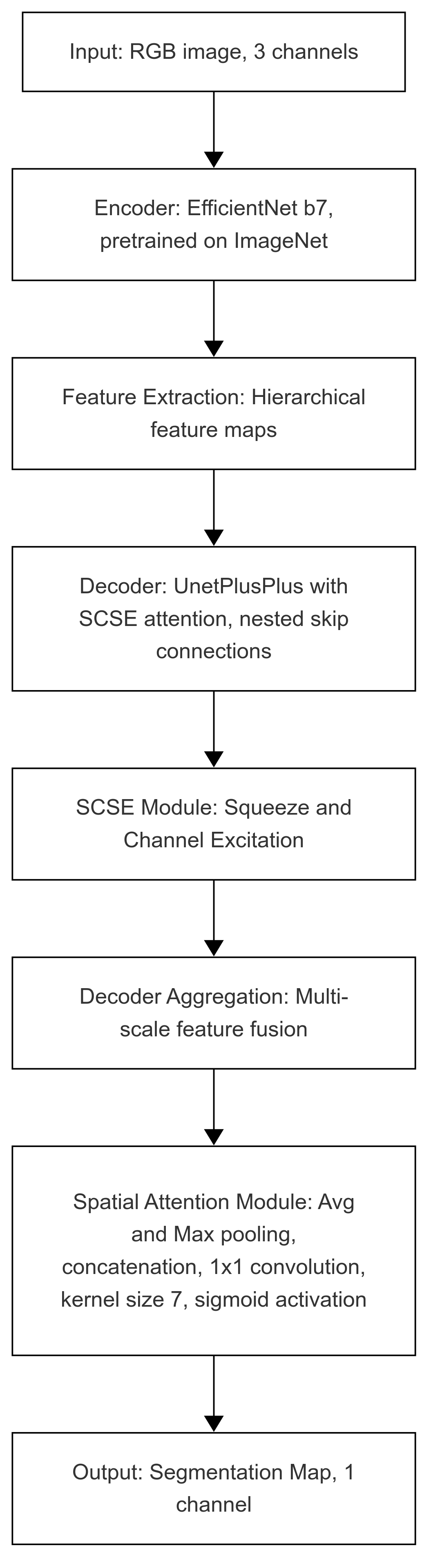}
    \caption{Overview of the segmentation pipeline: preprocessing, independent training of dual-attention U-Net++ models, ensemble predictions with TTA, and optional CRF-based refinement.}
    \label{fig:framework}
\end{figure}

To clearly illustrate our approach, we present an overview of our segmentation pipeline in Fig.~\ref{fig:framework}. The pipeline begins with preprocessing and intensive data augmentation, followed by the independent training of class-specific dual-attention U-Net++ models with EfficientNet-B7 encoders. Finally, ensemble predictions are generated using Test Time Augmentation and optionally refined via Conditional Random Fields post-processing.

\subsection{Dual-Attention U-Net++ Architecture}

\begin{figure}[H]
    \centering
    \includegraphics[width=0.7\textwidth]{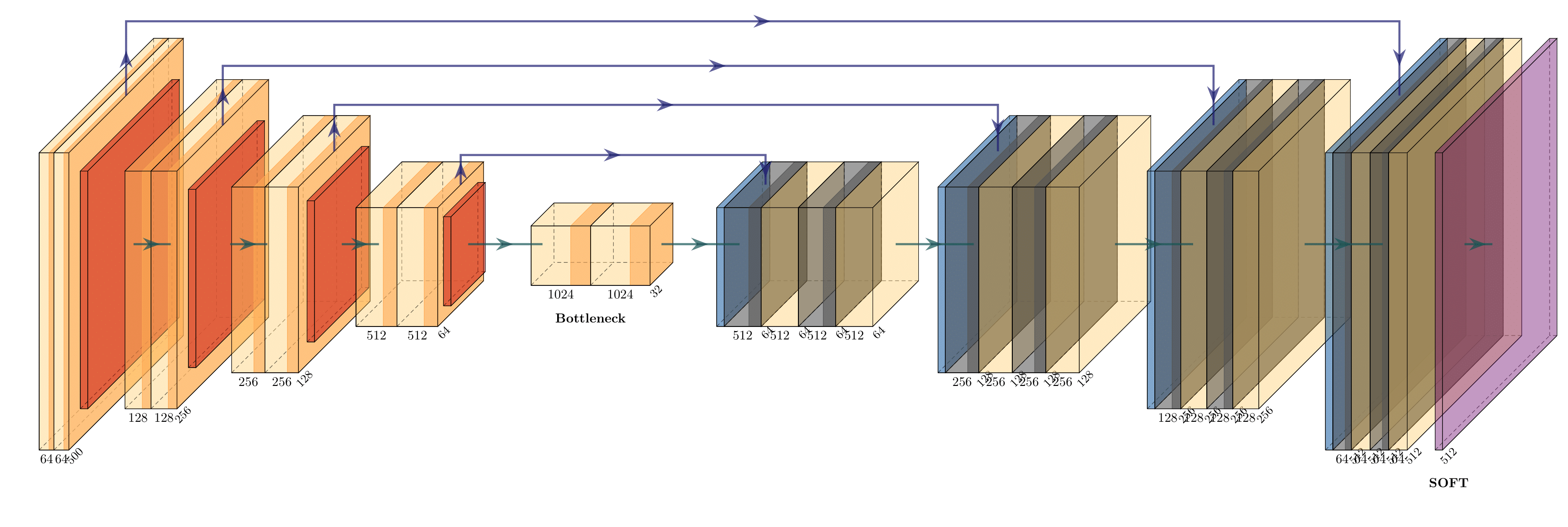}
    \caption{Schematic visualization of the dual-attention U-Net++ architecture with EfficientNet-B7 encoder. Attention modules (Spatial + SCSE) recalibrate feature maps, emphasizing relevant spatial regions and discriminative channels.}
    \label{fig:architecture}
\end{figure}

A detailed visualization of the dual-attention U-Net++ architecture is provided in Fig.~\ref{fig:architecture}. The model uses EfficientNet-B7 as the encoder backbone to extract high-level image features. Nested skip-connections characteristic of U-Net++ facilitate efficient feature reuse. Dual-attention modules, combining spatial and SCSE (channel-wise) attention mechanisms, are integrated within the decoding pathway to enhance segmentation accuracy.

\section{Results}\label{sec:results}

Our proposed dual-attention U-Net++ architecture, optimized via Bayesian sweeps, was evaluated on the NBC 2025 \& PCBBE 2025 segmentation challenge. Due to dataset imbalance, we trained separate models for wounds and scale markers, achieving robust internal validation scores as summarized in Table~\ref{tab:validation_results}.

\begin{table}[h!]
    \centering
    \small
    \begin{tabular}{|c|c|c|}
        \hline
        \textbf{Run} & \textbf{Wound segmentation (Dice)} & \textbf{Scale marker segmentation (Dice)} \\
        \hline
        1 & 0.9693 & 0.9028 \\
        2 & \textbf{0.9743} & \textbf{0.9281} \\
        3 & 0.9733 & 0.9175 \\
        4 & 0.9705 & 0.9258 \\
        \hline
        \textbf{Average} & 0.9718 & 0.9185 \\
        \hline
    \end{tabular}
    \caption{Internal validation Dice scores for wounds and scale markers across four different training runs. Best results for each class are shown in bold.}
    \label{tab:validation_results}
\end{table}

Final submissions were based on an ensemble of two scale marker models combined with a single, robust wound segmentation model.

The primary evaluation metric, chosen by the organizers, was a weighted F1-score (75\% wound segmentation, 25\% scale marker segmentation), computed externally on undisclosed hardware to ensure fairness. At the time of manuscript preparation, our models' official evaluation by competition organizers is ongoing; preliminary internal tests suggest high reliability and robustness.

Our final externally validated weighted F1-score was 0.8640.

\section{Discussion}\label{sec:discussion}

Segmenting wounds and scale markers poses significant challenges, including class imbalance, variable imaging conditions, and complex boundaries. Our approach addressed these issues through specialized training pipelines, dual-attention mechanisms, and Bayesian hyperparameter tuning.

\textbf{Comparison with Similar Studies.} Similar research efforts, such as those by Niri et al. \cite{niri2025wound} and Zimmermann et al. \cite{zimmermann2023wound}, explored attention-enhanced U-Net architectures. Niri et al. reported impressive Dice scores (approximately 94.1\%) with a dual-attention VGG16-based U-Net architecture. Our study extends these prior efforts by leveraging the powerful EfficientNet-B7 backbone, rigorous Bayesian hyperparameter optimization, and dedicated class-specific models. Unlike previous single-model multi-class approaches, our method facilitated targeted optimizations to handle severe class imbalance more effectively.

\textbf{Clarification on EfficientNet Encoder Selection.} While our initial benchmarking results (Fig.~\ref{fig:encoders}, Fig.~\ref{fig:architectures}) demonstrated EfficientNet-B0 as the top-performing encoder, we chose EfficientNet-B7 due to its architectural similarity yet greater representational capacity. Previous studies consistently highlight that higher-capacity EfficientNet variants significantly enhance generalization and robustness, which aligns with our objective of achieving superior segmentation performance in complex and highly variable medical imaging contexts.

Initially, extensive benchmarking identified EfficientNet-B7 as optimal despite its computational demands, justified by significant performance gains in validation Dice scores. The dual-attention enhancement of U-Net++ provided significant improvements in capturing spatial and channel-wise context.

Independent training for each class proved essential in managing dataset imbalance, aligning with contemporary research \cite{info15030140}. Furthermore, adaptive loss weighting strategies demonstrated clear advantages, although occasionally causing unstable training, including sudden learning rate spikes leading to training collapse (Dice dropping to 0), as visualized in Fig.~\ref{fig:class2}.

Test Time Augmentation consistently improved segmentation accuracy, emphasizing its role in real-world medical imaging. Future work could explore transformer-based architectures and more sophisticated ensemble strategies to improve accuracy further.

\section{Conclusion}\label{sec:conclusion}

In this study, we developed a robust and effective segmentation framework based on a dual-attention enhanced U-Net++ architecture for precise wound and scale marker segmentation. Our systematic benchmarking across various neural network architectures and encoder backbones identified EfficientNet-B7 as the optimal encoder, significantly enhancing segmentation accuracy and generalization capabilities. To address the pronounced class imbalance, we implemented class-specific training strategies, including tailored preprocessing, extensive data augmentation, and Bayesian hyperparameter optimization. These measures were critical in stabilizing model performance and achieving superior segmentation quality.

The incorporation of dual-attention mechanisms—combining spatial and channel-wise (SCSE) attention modules—proved essential for capturing complex features inherent in medical imagery, thereby improving both localization accuracy and boundary precision. Moreover, the use of ensemble prediction strategies combined with Test Time Augmentation further enhanced the reliability and robustness of our approach.

Evaluated externally on a challenging benchmark provided by the NBC 2025 \& PCBBE 2025 competition, our proposed method achieved a weighted F1-score of 0.8640, demonstrating significant potential for practical, clinical adoption in wound management workflows. These results underscore the benefits of advanced architectural enhancements, rigorous hyperparameter tuning, and ensemble methodologies for tackling complex, imbalanced medical image segmentation tasks.

Future work will explore the integration of transformer-based models, more sophisticated post-processing techniques, and the evaluation of the framework on diverse clinical datasets to further improve generalization and reliability, bringing automated, accurate wound assessment closer to routine clinical practice.

\section*{Declarations}
\begin{itemize}
\item \textbf{Funding}: Not applicable.
\item \textbf{Conflict of interest}: Authors declare no conflicts of interest.
\item \textbf{Ethics approval and consent to participate}: Not applicable.
\item \textbf{Consent for publication}: Not applicable.
\item \textbf{Data availability}: Dataset provided by NBC2025 Challenge organizers.
\item \textbf{Materials availability}: Not applicable.
\item \textbf{Code availability}: Available upon reasonable request.
\item \textbf{Author contribution}: Conceptualization and methodology: DC, JR, MR, OO; Software, validation, original draft preparation: DC; Review \& editing: MR, OO, JR.
\end{itemize}

\bibliography{references}

\begin{appendices}

\section{Additional Training Details}

All models were implemented using PyTorch Lightning and trained on NVIDIA A5000 GPUs using mixed precision. The final ensemble of scale-marker segmentation models was implemented as follows:

\begin{algorithm}[ht!]
\caption{Ensemble Prediction with CRF Postprocessing}\label{alg:ensemble}
\begin{algorithmic}[1]
\Require Image $I$, Trained models $M_1, M_2$, weights $w_1, w_2$, CRF iterations
\State $\text{prob1} \gets \text{sigmoid}(M_1(I))$
\State $\text{prob2} \gets \text{sigmoid}(M_2(I))$
\State $\text{ensemble\_prob} \gets w_1 \cdot \text{prob1} + w_2 \cdot \text{prob2}$
\State $\text{binary\_mask} \gets \text{ensemble\_prob} > 0.5$
\State $\text{final\_mask} \gets \text{CRF\_postprocess}(I, \text{binary\_mask}, \text{iterations})$
\State \Return $\text{final\_mask}$
\end{algorithmic}
\end{algorithm}

\begin{figure}[ht!]
    \centering
    \includegraphics[width=0.9\textwidth]{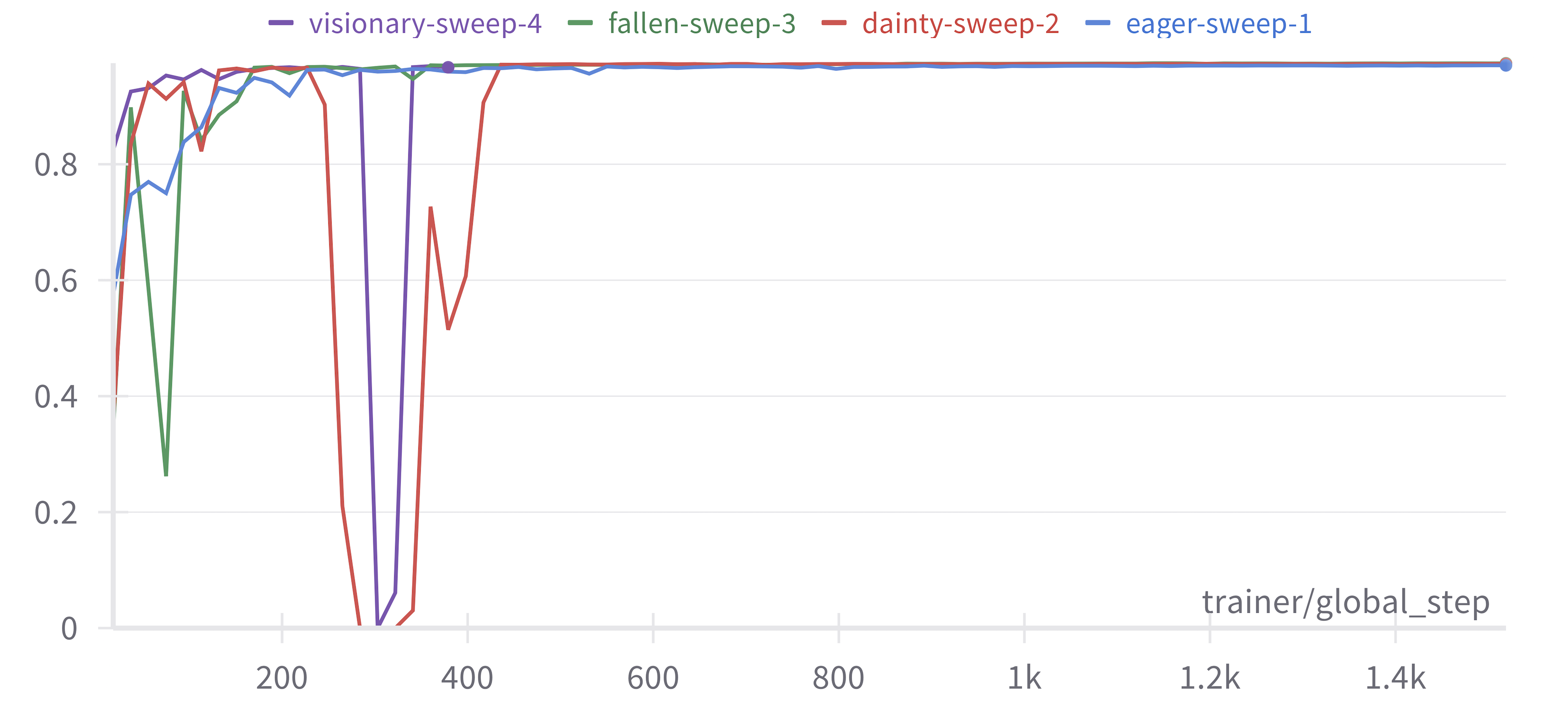}
    \caption{Training validation Dice scores for class 1 (wounds). Model demonstrated stable convergence around 97\%.}
    \label{fig:class1}
\end{figure}

\begin{figure}[ht!]
    \centering
    \includegraphics[width=0.9\textwidth]{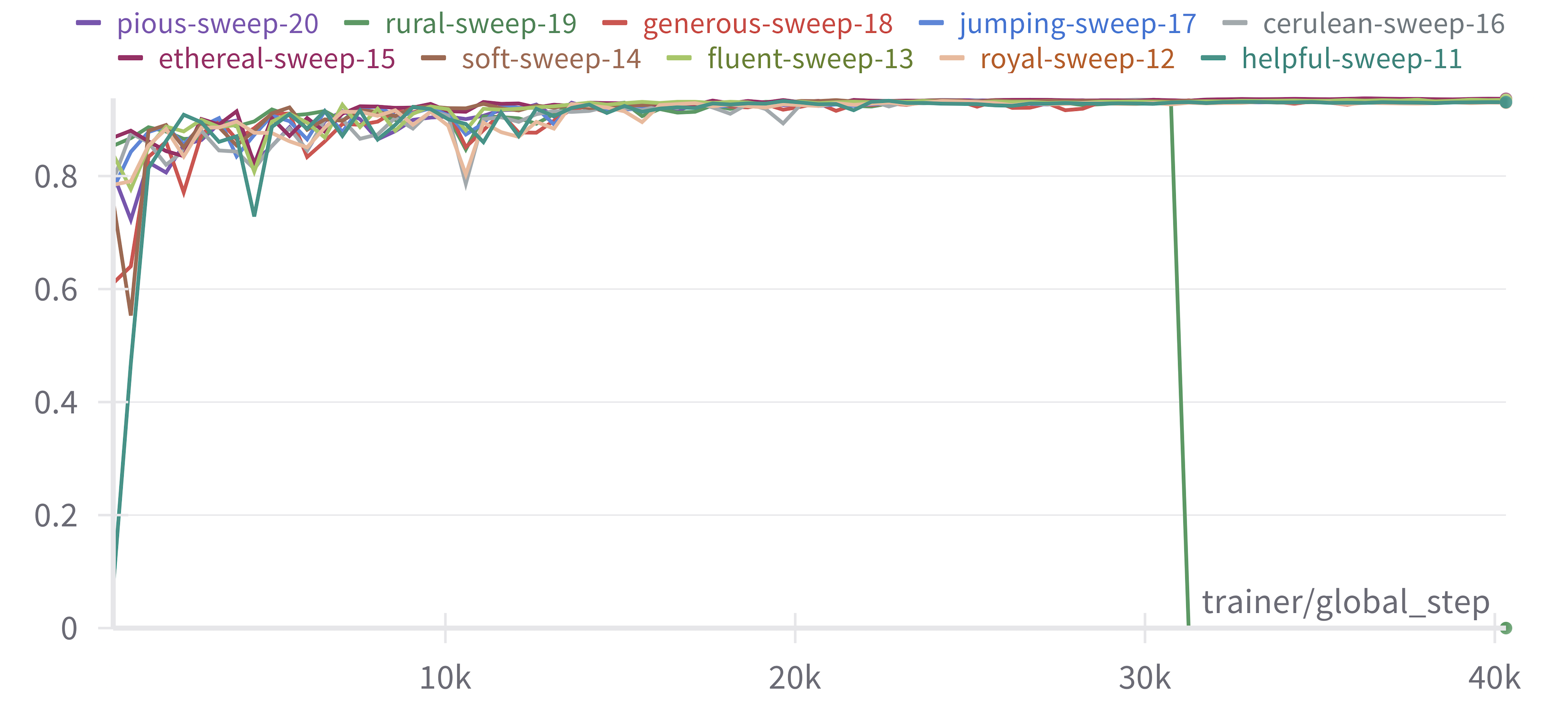}
    \caption{Training validation Dice scores for class 2 (scale markers). Note the instability due to adaptive loss functions, including a sudden collapse (Dice $\approx 0$) during training.}
    \label{fig:class2}
\end{figure}

\section{Hyperparameter Optimization}

Hyperparameter optimization was performed using Bayesian sweeps (Weights \& Biases). Key optimized parameters included:
\begin{itemize}
    \item Loss ratios (BCE, Dice, Focal)
    \item Mixup alpha (optimized: 0.3604)
    \item Augmentation strength (optimized: 1.137)
\end{itemize}

Sweep configurations and resulting optimal hyperparameters are documented and available in the project's WandB repository.

\end{appendices}

\end{document}